\pdfoutput=1

\documentclass[aps,11pt,onecolumn,superscriptaddress]{revtex4}
\usepackage{amsmath}
\usepackage{latexsym}
\usepackage{amssymb}
\usepackage{graphics,epstopdf}

\begin{document}

\title{Optimal free-will on one side in reproducing the singlet correlation}

\author{Manik Banik}
\email{manik_r@isical.ac.in}
\affiliation{Physics and Applied Mathematics Unit, Indian Statistical Institute, 203 B.T. Road, Kolkata-700108, India}

\author{MD Rajjak Gazi}
\email{rajjak_r@isical.ac.in}
\affiliation{Physics and Applied Mathematics Unit, Indian Statistical Institute, 203 B.T. Road, Kolkata-700108, India}

\author{Subhadipa Das}
\email{sbhdpa.das@bose.res.in}
\affiliation{S.N. Bose National Center for Basic Sciences, Block JD, Sector III, Salt Lake, Kolkata-700098, India}

\author{Ashutosh Rai}
\email{arai@bose.res.in}
\affiliation{S.N. Bose National Center for Basic Sciences, Block JD, Sector III, Salt Lake, Kolkata-700098, India}

\author{Samir Kunkri}
\email{skunkri@yahoo.com}
\affiliation{Mahadevananda Mahavidyalaya, Monirampore, Barrakpore, North 24 Parganas-700120, India}

\begin{abstract}

Bell's theorem teaches us that there are quantum correlations that can not be simulated by just shared randomness (Local Hidden variable). There are some recent results which simulate singlet correlation by using either 1 cbit or a binary (no-signaling) correlation which violate Bell's inequality maximally. But there is one more possible way to simulate quantum correlation by relaxing the condition of independency of measurement on shared randomness. Recently, MJW Hall showed that the statistics of singlet state can be generated by sacrificing measurement independence where underlying distribution of hidden variables depend on measurement direction of both parties [Phys. Rev. Lett.{\bf105} 250404 (2010)]. He also proved that for any model of singlet correlation, $86\%$ measurement independence is optimal. In this paper, we show that $59\% $ measurement independence is optimal for simulating singlet correlation
when the underlying distribution of hidden variables depend only on measurements of one party. We also show that a distribution corresponding to this optimal lack of free will, already exists in literature which first appeared in the context of detection efficiency loophole.

\end{abstract}

\maketitle 
\section{Introduction}

It has been found that the predictions made from quantum mechanical correlation violate some statistical inequalities based on
twin assumptions of locality and reality, these are
commonly known as Bell-type inequalities \cite{bell}. The possible (non-quantum) resource to reproduce the quantum mechanical correlations are the following:\\
(i)  Classical communication, aided by shared randomness,\\
(ii) Some no-signaling non-local correlation (P-R box\cite{pr}) aided by shared randomness,\\
(iii) Lack of free will on the part of parties performing local measurements\cite{hg, bg}.

Toner-Bacon \cite{bt} have shown that quantum correlation
of singlet state can be simulated by $1$ c-bit communication (signaling correlation) where measurement output is deterministic.
Cerf et.al. \cite{ce} have also reproduced the correlation of
singlet state by using binary input output no-signaling correlation and in this case measurement output is completely random.
Recently, Kar \cite{kar} and Hall \cite{hm}
independently conjectured  a trade off relation between amount of signaling correlation (classical communication)
and indeterminism for output in simulating singlet correlation.
In all the above models, the assumption of measurement independence has been kept intact.
Later in \cite{hg}, Hall showed that relaxation of measurement
independence can be a useful resource for modeling singlet state correlation.
He provided a deterministic model of singlet state in which only $14\%$ measurement
dependency is required.

In this paper, following the approach in \cite{hm},  we address the question how much measurement dependency is required to model the singlet state correlation and P-R correlation \cite{pr}
if only one party gives up his/her experimental free will while other party enjoys full experimental free will.
 Here we have shown that $41\%$ measurement dependency on one side is necessary
for simulating the singlet correlation. Interestingly there is a model which explicitly simulates the singlet correlation
showing that $41\%$ measurement dependency is also sufficient. Finally, by constructing some toy model, we discuss the relation between lack of free will and amount of violation of Bell-CHSH inequality.

\section{Lack of free will and modified Bell inequality }
Consider a set of statistical correlations $\{p(a,b|X,Y)\}$ of binary input-output system where $X$ and $a$ are respectively
input and output on one side and similarly $Y$ and $b$ for other side. If there is an underlying variable $\lambda$ in such model which reproduces this correlation, then
\begin{equation} \label{underlying}
p(a,b|X,Y) = \int d\lambda \rho(\lambda|XY) p(a,b|X,Y,\lambda)
\end{equation}
Obviously when  $\rho(\lambda|XY) =  \rho(\lambda)$, there is no lack of free will. Then the degree of measurement dependency is simply quantified by
the variational distance between the distribution of the shared parameter for any pair of measurement settings \cite{hg,hw}.
\begin{equation} \label{mdep}
M:= \sup_{X,X',Y,Y'} \int d\lambda \,\left|\rho(\lambda|XY) - \rho(\lambda|X'Y')\right|.
\end{equation}
The measurement independence is the property that the distribution of the underlying variable is independent of
 the measurement settings, i.e.,
\begin{equation} \label{free}
\rho(\lambda|X,Y) = \rho(\lambda|X',Y')
\end{equation}
for any joint settings $(X,Y)$, $(X',Y')$.
The degree to which an underlying model violates measurement independence is most simply quantified by the variational distance \cite{hg,hw}.

We now derive the required Bell's inequality. Let $X,X'$ and $Y,Y'$ denote possible measurement settings for a first and second observers, respectively,
and label each measurement outcome by $1$ or $-1$. Let $\langle XY\rangle$ denote the average of the product of the measurement outcomes,
for joint measurement settings $X$ and $Y$. If for any correlation model for which the assumption of no communication,
determinism and complete free will hold, the above statistics
satisfy the following so called Bell-CHSH \cite{chsh} inequality
\begin{equation} \label{bell}
 |\langle XY\rangle + \langle XY'\rangle + \langle X'Y\rangle - \langle X'Y'\rangle|   \leq 2
\end{equation}
By invoking measurement dependency, we now derive the following result.\\
\emph{Any underlying deterministic no-signaling model characterized by measurement dependence $M$  for one side would satisfy}
\begin{equation}
\label{nonlocal}B= |\langle XY\rangle + \langle XY'\rangle + \langle X'Y\rangle - \langle X'Y'\rangle | \leq 2 + M ~~\mbox{where}~~0\leq M\leq 2
\end{equation}
{\bf{Derivation}}---A deterministic hidden variable model in which distribution of hidden variables depend on the choice of measurement settings, the Bell-CHSH expression can be written as:
\begin{eqnarray}
B=|\langle XY\rangle + \langle XY'\rangle + \langle X'Y\rangle - \langle X'Y'\rangle|\nonumber \\=
|\int [\rho(\lambda|XY) V(X|\lambda) V(Y|\lambda)\nonumber
+\rho(\lambda|XY') V(X|\lambda) V(Y'|\lambda)\nonumber +\\
\rho(\lambda|X'Y) V(X'|\lambda) V(Y|\lambda)
 -\rho(\lambda|X'Y') V(X'|\lambda) V(Y'|\lambda] d\lambda|
\end{eqnarray}
Since the model is deterministic then for any given $\lambda$ each observable takes a definite value i.e
$V(X|\lambda),V(X'|\lambda),V(Y|\lambda),V(Y'|\lambda)$ attain a fixed value from the set $\{+1, -1\}$.

Now, for one sided measurement dependency models, without loss of generality, we assume that the distribution of $\lambda$ depends only on Alice's side measurements $X$ and $X'$.
Then, this condition can be expressed as
\begin{eqnarray}
\rho(\lambda|XY)=\rho(\lambda|XY')=\rho(\lambda|X) \nonumber \\ 
\rho(\lambda|X'Y)=\rho(\lambda|X'Y')=\rho(\lambda|X')\nonumber
\end{eqnarray}
Under the above two conditions the Eq.(6) now reduces to
\begin{eqnarray}
B=|\langle XY\rangle + \langle XY'\rangle + \langle X'Y\rangle - \langle X'Y'\rangle|\nonumber \\=
|\int [\rho(\lambda|X) V(X|\lambda) V(Y|\lambda)\nonumber
+\rho(\lambda|X) V(X|\lambda) V(Y'|\lambda)\nonumber +\\
\rho(\lambda|X') V(X'|\lambda) V(Y|\lambda)
 -\rho(\lambda|X') V(X'|\lambda) V(Y'|\lambda] d\lambda|
\end{eqnarray}
Next, since absolute value of an integral is less than or equal to integral of absolute value of the integrand ($|\int f(x)dx|\leq \int|f(x)|dx$), from Eq.(7) we obtain 
\begin{eqnarray}
B\leq
\int|\rho(\lambda|X) V(X|\lambda) V(Y|\lambda)\nonumber
+\rho(\lambda|X) V(X|\lambda) V(Y'|\lambda)\nonumber \\
~~~~~~~~~~~~~+ \rho(\lambda|X') V(X'|\lambda) V(Y|\lambda)
 -\rho(\lambda|X') V(X'|\lambda) V(Y'|\lambda) |~d\lambda \nonumber \\
 =\int |V(Y|\lambda)\left\{\rho(\lambda|X) V(X|\lambda)+\rho(\lambda|X') V(X'|\lambda)\right\} \nonumber \\
 ~~~~~~~~+ V(Y'|\lambda)\left\{\rho(\lambda|X) V(X|\lambda)-\rho(\lambda|X') V(X'|\lambda)\right\}|~d\lambda
\end{eqnarray}
Now, applying the triangle inequality in the integrand we get
\begin{eqnarray}
B \leq \int [~|~V(Y|\lambda)\left\{\rho(\lambda|X) V(X|\lambda)+\rho(\lambda|X') V(X'|\lambda)\right\}| \nonumber \\
 ~~~~~~~~+|~V(Y'|\lambda)\left\{\rho(\lambda|X) V(X|\lambda)-\rho(\lambda|X') V(X'|\lambda)\right\}|~]~d\lambda \nonumber \\
 = \int [~|~V(X|\lambda)V(Y|\lambda)\left\{\rho(\lambda|X) +\rho(\lambda|X') \frac{V(X'|\lambda)}{V(X|\lambda)}\right\}| \nonumber \\
 ~~~~~~~~+|~V(X|\lambda)V(Y'|\lambda)\left\{\rho(\lambda|X) -\rho(\lambda|X') \frac{V(X'|\lambda)}{V(X|\lambda)}\right\}|~]~d\lambda
\end{eqnarray}
Since Alice's and Bob' observable attains a value $\pm 1$ we get,

\begin{eqnarray}
B \leq \int [~|~\left\{\rho(\lambda|X) + k ~\rho(\lambda|X') \right\}| 
 +|~\left\{\rho(\lambda|X) - k ~\rho(\lambda|X') \right\}|~]~d\lambda
\end{eqnarray}
where $k=\frac{V(X'|\lambda)}{V(X|\lambda)}$ which for a given $\lambda$ either takes value $+1$ or $-1$.
Then, the symmetry of the integrand on changing the value of $k$ implies that
\begin{eqnarray}
B \leq \int |\rho(\lambda|X) + \rho(\lambda|X')|~d\lambda +\int |\rho(\lambda|X) - \rho(\lambda|X')|~d\lambda \nonumber \\
 = 2 + \int |\rho(\lambda|X) - \rho(\lambda|X')|~d\lambda
\end{eqnarray}
Taking the supremum over Alice's measurements $X$ and $X'$ gives

\begin{equation}
B \leq 2 + \sup_{X,X'} \int |\rho(\lambda|X) - \rho(\lambda|X')|~d\lambda = 2 + M
\end{equation}

For other inequivalent forms of Bell-CHSH expression also one can easily obtain that $B\leq2+M$.
In a no-signalling deterministic model, the one-sided degree of measurement dependence that is required to reproduce the correlation corresponding to  $2\sqrt{2}$, is given by
\begin{eqnarray}
M&=&2(\sqrt{2}-1)\nonumber
\end{eqnarray}
The experimental free will is defined as\cite{hg}:
\begin{equation}
F=1-M/2
\end{equation}
For $M=0$, $F = 1$ implying no lack of free will (see eqn.(3)).
The experimental free will for $B = 2\sqrt{2}$, is given by
\begin{eqnarray}
F&=&1-M/2 =2-\sqrt{2}=.59
\end{eqnarray}
For modeling Bell- violation $4$ by a nosignaling deterministic model, we need measurement dependence
\begin{eqnarray}
&M&=2 \nonumber \\
&\Rightarrow& F= 0\nonumber
\end{eqnarray}
So in this case there is no experimental free will for Bob.\\
Now, we provide toy model with tabular form which reproduces above correlation and is also
compatible with the above results. The basic assumptions of these models are no-signalling and determinism.
\subsection{Table: Deterministic no-signaling model for correlations violating BI.}
\begin{center}
\begin{tabular}{lllllllll}
$\lambda$ &$X(\lambda)$ & $ X'(\lambda)$ &$Y(\lambda)$ & $Y'(\lambda)$& $\rho(\lambda|XY)$& $\rho(\lambda|XY')$ & $\rho(\lambda|X'Y)$ & $\rho(\lambda|X'Y')$\\
$\lambda_1$ & $-a$& $-a$& $-a$ & $a$ &$0$ &$0$ & $p$ & $p$\\
$\lambda_2$& $b$& $b$ & $b$ & $b$ & $1$ & $1$ &$1-p$ & $1-p$\\

\end{tabular}
\end{center}
In this Table, we describe the singlet model which contains two underlying variables $\lambda_1$ and $\lambda_2$ with
the outcome for measurement setting X denoted by X($\lambda_j$). For this model outcomes are specified by two numbers
$a, b \in \{1,-1\}$. Probability distributions for this model, are defined by single parameter,
$0\leq p \leq 1$, as per table. From this table, $ \langle XY\rangle = \langle X'Y\rangle = \langle XY'\rangle=1$ and
$ \langle X'Y'\rangle =1-2p$. Therefore, the Bell expression for this is $B=\langle XY\rangle + \langle XY'\rangle +
\langle X'Y\rangle - \langle X'Y'\rangle =2+2p$ and Bell-CHSH violation amount is $2p$ .
The degree of measurement dependency is $M=2p$ and hence clearly $0\leq M \leq 2$ and  $0\leq F \leq 1$.
Therefore, for $B=2+M$, the amount of one-sided free-will is $F=1-p$ in agreement with Eq.(10). From this table one can also see that maximal Bell violation (PR correlation) can be simulated only by sacrificing full free will of one party $(p=1)$; Interestingly, this can be achieved by sacrificing free will on both sides by $34\%$ \cite{hg}.

\section{Lack of free will of one party and  Singlet simulation}
For a binary input-output system, let $X$ and $Y$ denote Alice's and Bob's input. If Alice and Bob share a random
variable $\lambda_s$ which is real three dimensional unit vector, distributed according to a distribution with probability density
\begin{equation}
\rho(\lambda_s|X,Y)=\rho(\lambda_s|X)=|X.\lambda_s/2\pi|
\end{equation} i.e, their underlying distribution depends on only one party's measurement direction. It has already been shown
 that the singlet correlation can be reproduced by sharing the above type of distribution \cite{gis}.
Now, for this type of distribution, the amount of measurement independence is
\begin{equation}
M:= \sup_{X,X'} \int d\lambda \,\left||X.\lambda/2\pi| - |X'.\lambda/2\pi|\right|
\end{equation}
The value of $M$ can be computed, without loss of generality, if we choose a reference frame such as $X= (1,0,0)$, $X'= (\cos\beta,\sin\beta,0)$ and
$\lambda= (\sin\theta \cos\phi,\sin\theta \sin\phi,\cos\theta)$,then
\begin{eqnarray*}
M:=\sup_\beta \frac{1}{4}\int^{2\pi}_{0} ||\cos\phi|-|\cos(\phi-\beta)|| d\phi =2\sqrt{2}-2
\end{eqnarray*}
\\
Where the supremum value is attained for a particular setting $X'=(0,1,0)$.

So, corresponding fraction of measurement independence is
\begin{equation}
 F= 2- \sqrt{2}
\end{equation}

Note that there is another measure for the correlation between hidden variable $\lambda$ and measurement settings of Alice ($X$) and Bob ($Y$) provided in the model given by Barrett and Gisin \cite{bg}. In this model the measurement dependency is quantified by mutual information $I(x,y:\lambda)$. For one sided measurement dependency where the distribution of hidden variables is given by Eq.(15), it has been shown that $I(x,y:\lambda)=I(x:\lambda)\approx 0.28 ~ \mbox{bits}$.
\section{Conclusion}
The main result of this paper is that one sided measurement dependency can be used as a useful resource for simulating correlations
that violate Bell-CHSH inequality. Like classical communication and no-signaling non-local resource,
 lack of free will of one or both the parties can be used as
nonlocal resource. We find the optimal measurement independence for the first case in simulating singlet statistics
and relate it to a distribution of hidden variables which is function of measurement direction of one party.

\begin{acknowledgments}
It is a pleasure to thank Guruprasad Kar for many stimulating discussions. SD and AR acknowledge support
from DST project SR/S2/PU-16/2007.
\end{acknowledgments}


\end{document}